\documentstyle[epsfig,12pt,aps]{revtex}
\begin{document}
\begin{titlepage}
\begin{center}
      {\Large \bf
         Evidence for the fourth $P_{11}$ resonance predicted by
         the constituent quark model}
\vspace*{10mm}\\
      Simon Capstick\footnote{capstick@scri.fsu.edu} \\
      {\em Supercomputer Computations Research  Institute, \\
      and Department of Physics, Florida State University, Tallahassee,
      FL \vspace*{0.7cm} 32306-4130} \\
      T.-S. H. Lee\footnote{lee@anlphy.phy.anl.gov} \\
      {\em Physics Division, Argonne National Laboratory, Argonne,
      Illinois 60439, \vspace*{0.7cm} USA} \\
      W. Roberts\footnote{wroberts@nsf.gov} \footnote{on leave from Department 
      of Physics, Old Dominion University,
      Norfolk, VA 23529} \\
      {\em National Science Foundation \\ 
      4201 Wilson Boulevard, Arlington, VA 22230
            \vspace*{0.7cm}} \\
      A. \v{S}varc\footnote{svarc@rudjer.irb.hr} \\
      {\em Rudjer Bo\v{s}kovi\'{c} Institute, Zagreb, Croatia}
\end{center}
      It is pointed out that the third of five low-lying $P_{11}$
      states predicted by a constituent quark model can be identified
      with the third of four states in a solution from a three-channel
      analysis by the Zagreb group. This is one of the so-called
      ``missing'' resonances, predicted at 1880 MeV. The fit of the
      Zagreb group to the $\pi N \rightarrow \eta N$ data is the
      crucial element in finding this fourth resonance in the $P_{11}$
      partial wave.
\end{titlepage}
      The  study  of  nucleon  resonances  ($N^*$)  involves  two steps.
      First, the positions and the decay widths of the $N^*$'s  must  be
      extracted  from  the  available  experimental  data of $\pi N$ and
      $\gamma N$  reactions.   This  step  is  usually  accomplished  by
      performing a partial wave analysis (PWA).  The parallel step is to
      develop a  theoretical  explanation  of  the  extracted  resonance
      parameters.   This  is most commonly pursued by developing various
      quark models.

      The PWA of $\pi N$ scattering has a long history.  A number of
      approaches have been developed with various levels of
      sophistication in implementing some theoretical constraints in
      order to offset the difficulties due to the lack of complete and
      accurate data.  The most elaborate early PWA analyses were
      performed mainly for $\pi N$ elastic scattering using either
      single-channel dispersion relations\cite{Hoe} or a
      multi-channel, multi-resonance, unitary model\cite{Cut}.  Recent
      PWA's \cite{VPI,Man,Sva98,Day} are based on some variations of
      these approaches, but using updated data sets.  These efforts
      have led to some revisions of the resonance parameters listed by
      Particle Data Group (PDG) \cite{PDG}, and have triggered debates
      on some resonance parameters.  In particular, the $S_{11}$ and
      $P_{11}$ resonances have been frequently discussed, and this has
      stimulated new experimental efforts in order to resolve the
      existing controversies.  In this paper, we focus on the $\pi N$
      partial wave $P_{11}$, and discuss how the quark-model
      predictions of Refs.~\cite{FC,SS,CI,CR1}, which contain more
      states than listed by the PDG, are consistent with a PWA
      analysis~\cite{Sva98,Bat95,Bat98} based on a multi-channel,
      multi-resonance, unitary model first developed by Cutkosky and
      collaborators~\cite{Cut}.

      In Refs.~\cite{Bat95,Bat98},  it  was  pointed  out  that  two  PWA
      solutions  can  be  found  within a three-channel ($\pi N, \eta N,
      \pi^2 N$) unitary model.  They differ mainly in  the
      number  of  $P_{11}$  resonances and their corresponding branching
      ratios.  The analysis has been repeated in Ref.~\cite{Sva98}  using
      an  improved $S_{11}$ amplitude~\cite{S11}, which is
      crucial in constraining the fit to the $\pi N \rightarrow \eta  N$
      cross  sections  near  threshold.   We  are again able to find two
      solutions, as  shown  in  Table  1.   The  resonance
      parameters  are  essentially  the same for the three-resonance and
      four-resonance solutions,  with  the  exception  of  the  $P_{11}$
      channel.   In  particular, the branching ratios to the $\eta N$ and
      $\pi^2 N$ channels for the second $P_{11}(1710$ MeV) state satisfy
      $x_{\eta N} \gg x_{\pi^2 N}$ for the three-resonance solution, but
      $x_{\pi^2 N} \gg x_{\eta N}$ for the four-resonance  solution.

      To further distinguish these two solutions, it is necessary to
      extend the present analysis by replacing the $\pi^2 N$ channel
      with an explicit treatment of the inelastic data in each of the
      channels $\pi\Delta$, $\rho N$ , $\sigma[(\pi\pi)_{l=0}] N$,
      etc. This highly non-trivial task, while beyond the scope of
      this investigation, was carried out in the analysis of Manley
      and Saleski~\cite{Man}. The resulting total $\pi^2 N$ branching
      ratio is, therefore, more strongly constrained by the data than
      that of the Zagreb analysis.  We therefore assume that the more
      acceptable solution of the Zagreb analysis is that which yields
      a $\pi^2 N$ branching ratio closer to that of Manley and
      Saleski.  This is the four-resonance solution listed in Table 1.
      We note here that the data of $\pi N \rightarrow \eta N$ are not
      included in Manley and Saleski's analysis, but are treated with
      great care in the Zagreb analysis.  When this data and the
      $x_{\pi^2 N}$ values from Manley and Saleski are put together,
      the four-resonance solution is strongly favored.  This example
      clearly demonstrates that some ``missing'' resonances are
      sensitive to particular channels, and can be discovered only
      when the data associated with those channels are included in the
      analysis.

      The   results   of   constituent   quark  models  are  useful  for
      interpreting the results shown in Table 1.  In particular,  models
      which  treat  the  three  light  quarks  as  symmetric predict the
      existence of several positive-parity excited baryon states in  the
      1700-2000  MeV region which have not previously been identified in
      PWA's.  In particular, in the $P_{11}$ partial wave in $\pi N$,  a
      model  which perturbs around the spectrum of two three-dimensional
      harmonic  oscillators  with  hyperfine  and   linear   confinement
      corrections~\cite{IK}  found  four  $P_{11}$ excited states, which
      are part of the $N=2$  band  of  positive-parity  excited  states.
      Prior to configuration mixing by the perturbations, two are radial
      excitations of  the  nucleon,  with  either  totally-symmetric  or
      mixed-symmetry  spatial  wavefunctions,  one  is  a  total orbital
      angular momentum $L=2$ state with quark-spin-${3\over 2}$, and the
      fourth  is  an  $L=1$  state  with quark-spin-${1\over 2}$.  After
      mixing, the two radial excitations are identified  with  the  two
      low-lying  PDG  states  $P_{11}(1440)$  and  $P_{11}(1710)$ on the
      basis of their perturbed masses.  Also, an analysis of the $\pi N$
      decay  amplitudes  of  these  states  using  a  decay  model where
      point-like pions are emitted directly  from  the  quarks~\cite{KI}
      showed  that  these  two states should have stronger amplitudes to
      couple to the $\pi N$ channel than the the remaining two `missing'
      states  in  the  $N=2$  oscillator  band.  The more massive of the
      latter has the smallest predicted $\pi N$ amplitude.

      These predictions for  masses  and  $\pi  N$  decay  branches  are
      essentially    confirmed    and    extended   in   the   work   of
      Refs.~\cite{FC,SS} and~\cite{CI,CR1}.  These  models  went  beyond
      perturbing  around the $N=2$ oscillator band in the description of
      the spectrum, and used a microscopic quark model of strong  decays
      (the $^3P_0$ model) which ascribes structure to the emitted meson.
      We  will  focus  here,  for  definiteness,   on   the   model   of
      Refs.~\cite{CI,CR1},  which  predicts  four  $P_{11}$ states below
      2000 MeV (at 1540, 1770, 1880, and 1975 MeV), and many  additional
      excited states with wavefunctions predominantly in the $N=4$ band,
      the lightest of which is at 2065 MeV.

      From Table~1 it can be seen that the first three model states at
      1540, 1770 and 1880 MeV correspond nicely to those of the
      four-resonance solution of Zagreb analysis, while the third
      model state at 1880 MeV can not be identified with any of the
      PDG resonance parameters in the first column.  The PDG
      parameters are based mainly on analyses which do not
      ``explicitly'' account for the $\pi N \rightarrow \eta N$
      reaction data.  From Table~1 we see that the third model state
      at 1880 MeV has a substantial predicted partial decay width to
      $\eta N$ channel, and hence should be more easily identified in the
      Zagreb analysis.  It is possible that this ``missing'' resonance
      could also be found by the multi-channel analysis of Manley and
      Saleski if the data of $\pi N \rightarrow \eta N$ are included.

      The fourth model state at 1975 MeV does not correspond to any
      PDG state, and is not found in the Zagreb solutions.  This is
      not surprising, since this state is predicted to have very weak
      decay widths for the $\pi N$ and $\eta N$ channels.  It is
      possible that this state is more sensitive to a particular
      channel of the $\pi^2 N$ continuum, like $\pi
      \Delta$~\cite{CR1}, and can only be identified in an analysis in
      which the data for that particular inelastic channel are
      included explicitly.  However, the quality of the $\pi^2 N$ data
      at $W\simeq 2000$ MeV is not good enough for an accurate
      determination of the partial cross section to each individual
      inelastic channel.  This is perhaps the reason why this state is
      also not found in the multi-channel analysis of Manley and
      Saleski (included in the PDG results).

      A substantial discrepancy is found for the branching ratios
      $x_\eta$ and $x_{\pi^2}$ of the highest mass resonance at $\sim
      2100$ MeV.  However, the branching ratios extracted from the PWA
      at such high energies are heavily dependent on the input data of
      $\pi N\rightarrow \pi\pi N$ and $\pi N \rightarrow \eta N$
      reactions in the PWA analyses, and the quark model branching
      ratios should be considered upper bounds as some channels have
      been omitted from Ref.~\cite{CR1}.  It is also likely that there
      are substantial corrections to the constituent quark model from
      baryon-meson loops, and possible excitations of the glue at
      higher masses in the $P_{11}$ partial wave~\cite{CP}, which are
      ignored here.  Therefore, one expects improved agreement between
      the models when the input becomes more constrained, and more is
      known about the structure of higher-mass states.
      
      The  results  of analyses such as the one being discussed here are
      also very important for helping to distinguish  between  different
      models  of  the  nucleon  and  its excitations.  For instance, the
      so-called diquark models predict fewer states  in  the  excitation
      spectrum~\cite{diquark},   because  there  are  fewer  degrees  of
      freedom in the models.  However, such models  still  predict  more
      states  than  observed  experimentally.   In particular, they also
      predict a missing $P_{11}$ state near the one identified here.  To
      the  best of our knowledge, the only model that lacks such a state
      is that of Ref.~\cite{kirchbach}.

      In conclusion, we identify the need for a fourth resonance in  the
      $P_{11}$  partial wave when using the current $\pi N$ and $\eta N$
      data base for a multi-channel,  multi-resonance  PWA~\cite{Sva98}.
      The  extracted  positions  and  branching ratios are in reasonable
      agreement  with  quark-model  predictions~\cite{CR1},  given   the
      insufficiently determined input.

      This work is supported by the Department of Energy under
      Contract DE-FG02-86ER40273 and the Florida State University
      Supercomputer Computations Research Institute which is partially
      funded by the Department of Energy through Contract
      DE-FC05-85ER25000 (S.C.); by the Department of Energy through
      contracts DE-AC05-84ER40150 and DE-FG05-94-ER40832, and by the
      National Science Foundation through award PHY 9457892 (W.R.); by
      the U.S. Department of Energy, Nuclear Physics Division, under
      Contract No. W-31-109-ENG-38 (T.-S.H.L.); and by the US-Croatia
      international program under Contract JF 221 (A.S.).  W.R. is
      also grateful for the support and hospitality of the
      Universit\'e Joseph Fourier, Grenoble, France, and the Institut
      des Sciences Nucl\'eaires, Grenoble, France.  A.\v{S}. is
      grateful for the support and hospitality of the Swedberg
      Laboratory, Uppsala University, Uppsala, Sweden, where he has
      been given a lot of useful advice and perfect working conditions
      to complete this work.  We are saddened to inform our
      collaborators that one member of our small group, Mijo
      Batini\'{c}, has, after a brave fight against a long illness,
      deceased in Spring 1998.
\newpage
\begin{table}  
\squeezetable\label{Table:1}
\caption{Resonance parameters of the
      phenomenological~\protect{\cite{Sva98}} and the
      quark~\protect{\cite{CI,CR1}} models. The first column gives the
      masses, widths, and pion-decay branching fractions from the
      latest PDG compilation~\protect{\cite{PDG}}.}
\begin{tabular}{cccccccccccccccc}
                  & \multicolumn{10}{c}{{\bf Zagreb group}} &
                  \multicolumn{5}{c}{{\bf quark model of}} \\
 & \multicolumn{10}{c}{{\bf Ref.~\protect{\cite{Sva98}}}} 
 & \multicolumn{5}{c}{{\bf Refs.~\protect{\cite{CI,CR1}}}}  \\
\hline
     States       & \multicolumn{5}{c}{Three P$_{11}$ resonances}    
    &  \multicolumn{5}{c}{Four P$_{11}$ resonances}  &
      \multicolumn{5}{c}{Five P$_{11}$ resonances} \\
\hline
      L$_{2I,2J}$ & Mass & Width & $x_\pi$ & $x_{\eta} $ & $x_{\pi^2} $ 
      & Mass & Width & $x_\pi$ & $x_{\eta} $ & $x_{\pi^2} $ & Mass & Width
      & $x_\pi$ & $x_{\eta} $ & $x_{\pi^2}$ \\
      ${\rm (_{Mass/Width}^{x_{\pi}})}$ & (MeV) & (MeV) & (\%) & (\%) 
      & (\%) & (MeV) & (MeV) & (\%) & (\%) & (\%) & (MeV) & (MeV) 
      & (\%) & (\%) & (\%) \\
 \hline \hline
      S$_{11}(_{1535/120}^{38})$   & 1552(16) & 181(12) & 45(7) & 51(6)
      & 4(4)   & 1553(8)  & 182(25) & 46(7)  & 50(7)        & 4(2)      
      & 1460     & 645     & 34    & 66         & 0 \\
      S$_{11}(_{1650/180}^{61})$   & 1653(12) & 205(18) & 76(6) & 19(7) 
      & 5(3)   & 1652(9)  & 202(16) & 79(6)  & 13(5)        & 8(3)
      & 1535     & 315     & 47    & 39         & 14 \\
      S$_{11}(_{2090/95 \:  }^{9})$& 1809(21) & 380(50) & 30(7) & 20(6)
      & 50(8)  & 1812(25) & 405(40) & 32(6)  & 22(10)       & 46(9)        
      & 1945     & 595     & 6     & 2          & 89 \\ \\
      P$_{11}(_{1440/135}^{51})$   & 1437(21) & 401(40) & 60(7) & 0(0)
      & 40(6)  & 1439(19) & 437(14) & 62(4)  & 0(0)         & 38(4)        
      & 1540     & 425     & 97    & 0          & 3 \\
      P$_{11}(_{1710/120}^{12})$   & 1713(25) & 160(20) & 20(5) & {\bf 78(3)}
      & 2(8)   & 1729(16) & 180(17) & 22(24) & 6(8)         &{\bf 72(23) }
      & 1770     & 305     & 6     & 22         & {\bf 72} \\
      P$_{11}$                     & -        &      -  &    -  &    - 
      &    -   & 1740(11) & 140(25) & 28(34) & 12(9)        &{\bf 60(35)}  
      & 1880     & 155     & 5     & 18         & {\bf 76} \\
      P$_{11}$                     & -        &      -  &    -  &    - 
      &    -   &    -     &    -    &   -    &   -          &     -        
      & 1975     & 45      & 8     & 0          & 92    \\
      P$_{11}(_{2100/200}^{ 9})$   & 2161(30) & 380(60) & 14(6) &{\bf 82(8)}
      &  4(6)  & 2157(42) & 355(88) & 16(5)  &{\bf 83(5)}   &  1(1)
      & 2065     & 270     & 22    & 1          & {\bf 77} \\ \\
      D$_{13}(_{1520/114}^{54})$   & 1522(8)  & 130(10) & 50(4) & 0.1(0.1)
      & 49(4)  & 1522( 8) & 132(11) & 55(5)  & 0.1(0.1)     & 45(5)
      & 1495     & 115     & 64    & 0          & 36 \\    
      D$_{13}(_{1700/110}^{ 8})$   & 1809(15) & 138(30) & 10(3) & 10(3)
      & 80(6)  & 1817(22) & 134(37) &  9(6)  &  14(5)       & 77(9)         
      & 1625     & 815     &  4    & 0          & 96 \\
      D$_{13}(_{2080/265}^{ 6})$   & 2001(16) & 610(50) & 15(8) & 6(2)
      & 79(7)  & 2048(65) & 529(13) & 17(7)  &   8(3)       & 75(7)        
      & 1960     & 535     & 12    & 6          & 81 \\
\end{tabular}
\end{table}


\begin{thebibliography}{Sva82a}
\bibitem{S11} The work of Refs.~\cite{Bat95,Bat98} was based on
           the KH80 solution of Ref.~\cite{Hoe}. The improvements to
           the $S_{11}$ amplitude are based on replacing the KH80
           solution below 1500 MeV in the $S_{11}$ partial wave with
           the SM95 solution from Ref.~\cite{VPI}. For details see
           Ref.~\cite{Sva98}.
\end{thebibliography}
\end{document}